\documentclass[twocolumn,twoside,slac_two]{revtex4}
\usepackage{graphicx}
\usepackage{fancyhdr}
\begin{document}

\title{VERITAS Observations of the Unidentified Point Source \\HESS J1943+213}

%

\author{K. Shahinyan for the VERITAS Collaboration}
\affiliation{University of Minnesota, Minnesota Institute for Astrophysics, Minneapolis, MN, 55455}

\begin{abstract}
The H.E.S.S. Galactic plane scan has revealed a large population of Galactic very high energy (VHE; E $\textgreater$ 100 GeV) emitters. The majority of the galactic sources are extended and can typically be associated with pulsar wind nebulae (35\%) and supernova remnants (21\%), while some of the sources remain unidentified (31\%)~\cite{carrigan}. A much smaller fraction of point-like sources (5 in total, corresponding to 4\%) are identified as gamma-ray binaries. Active galactic nuclei located behind the Galactic plane are also a potential source class. An active galaxy could be identified in the VHE regime by a point-like appearance, a high variability amplitude (up to a factor of 100) and a typically soft spectrum (due to absorption by the extra-galactic background light). Here we report on VERITAS observations of HESS J1943+213, an unidentified point source discovered to emit above 470 GeV during the extended H.E.S.S. Galactic plane scan~\cite{hess}. This source is thought to be a distant BL Lac object behind the Galactic plane and, though it exhibits a steep spectrum it is a weak GeV source, only recently detected using 5 years of Fermi-LAT data~\cite{peter}. Deep VERITAS observations at high elevations result in the most significant VHE detection of this object so far, with an excess above 200 GeV of more than 18$\sigma$. We use variability and spectral analyses of VERITAS data on HESS J1943+213 in a multi-wavelength context to address the source classification.

\end{abstract}

\maketitle

\thispagestyle{fancy}


\section{Identity of HESS J1943+213}
HESS J1943+213 was first discovered in very-high-energy (VHE; E $\textgreater$ 100~GeV) gamma rays during 
the H.E.S.S. Galactic plane scan~\cite{hess}. Due to its point-like appearance in VHE gamma rays, three possible
source classes were suggested: gamma-ray binary, pulsar wind nebula (PWN), and BL Lac object (blazar). 

Assuming the source is a gamma-ray binary, ref.~\cite{hess} used the lack of detection of a massive (O- or B-type) companion star to estimate a distance limit of greater than $\sim$25 kpc. This distance would place the binary well beyond the extent of the Galactic disk and would imply an X-ray luminosity 100-1000 times higher than known gamma-ray binaries. Such a distance limit is therefore problematic, and this scenario is disfavored. In addition, the point-like appearance in the X-rays and the soft VHE spectrum with a power-law index of $\Gamma$=3.1 $\pm$ 0.3 (in contrast to the softest known PWN index of 2.7) motivated ref.~\cite{hess} to argue against the PWN scenario, leaving the blazar hypothesis. The authors found all observations to be consistent with the blazar scenario. The point-like nature in both X-rays and VHE, the soft VHE spectral index, and a preliminary IR spectrum showing lack of emission lines are expected for a blazar. Moreover, the hard X-rays observed with INTEGRAL IBIS and Swift BAT instruments show no evidence of a cutoff up to an energy of $\sim$195~keV. If the source is a blazar, it would be categorized as an extreme high-synchrotron-peak BL Lac object (extreme HBL), a class of blazars with the synchrotron peak located at energies \textgreater1 keV~\cite{ehbl}.

Since the discovery publication, the identity of HESS J1943+213 has been the topic of an ongoing debate. 1.6-GHz VLBI observations of the HESS J1943+213 counterpart with the European VLBI Network produced
a detection that was claimed to show extension, with FWHM angular size of 15.7~mas (the expected size for a point source is 3.5~mas)~\cite{gabanyi}. Based on this measurement, the brightness temperature of the counterpart was estimated to be 7.7$\times$10$^{7}$ K and was used to argue against the blazar scenario, as the expected brightness temperature of HBLs is in the 10$^{8}$--10$^{9}$ K range. In addition, ref.~\cite{gabanyi} employed a 1$'$ feature observed in the 1.4-GHz VLA C-array configuration image to support the PWN hypothesis, with the assertion that the angular size of the feature is consistent with a Crab-like PWN placed at a distance of 17~kpc. On the other hand, ref.~\cite{tanaka} argued in favor of an extreme HBL by constructing a spectral energy distribution and drawing comparisons to a known extreme HBL, 1ES 0347-121. More recently, ref.~\cite{peter} bolstered the extreme HBL case by observing the near-infrared (K-band) counterpart of HESS J1943+213, claiming potential detection of an elliptical host galaxy with 10\% probability the object is a star. Using 5 years of data, they also obtain the first significant detection of the source with Fermi LAT (within 2$\sigma$ positional uncertainty) in the 1--300 GeV energy regime. Nonetheless, there is yet to be a definitive identification of the HESS J1943+213 source class.

Here we present results from VERITAS observations of HESS J1943+213 and discuss implications for the debate regarding the source identity.

\begin{figure}[h]
\includegraphics[width=85mm]{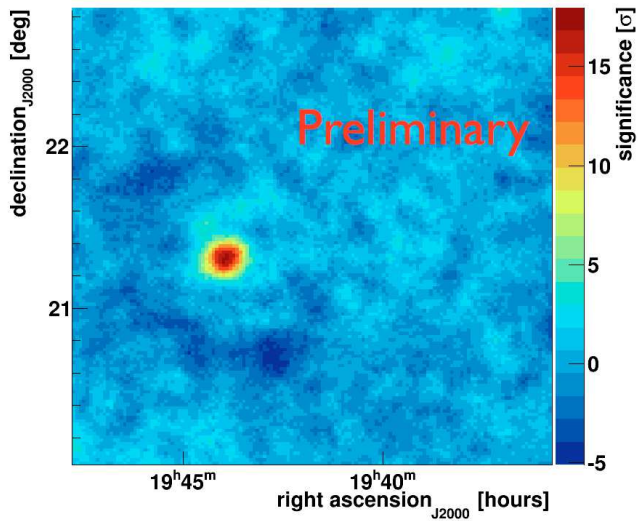}
\caption{Significance sky map of VERITAS observations of HESS J1943+213.}
\label{sigmap}
\end{figure}

\section{VERITAS Observations}
Very Energetic Radiation Telescope Array System (VERITAS) is an array of four 12-m imaging 
atmospheric Cherenkov telescopes, located in Arizona, USA at an elevation of $\sim$1270~m. The camera for 
each telescope is composed from 499 photo-multiplier tubes, with a field of view of approximately 3.5$^{\circ}$.
VERITAS is able to reliably reconstruct VHE gamma rays with energies between 100~GeV and 30~TeV~\cite{holder}, with the
sensitivity to detect a 1\% Crab Nebula flux source at 5$\sigma$ in 25 hours. The systematic uncertainty in the energy 
determination is within the 15--20\% range.

VERITAS observed HESS J1943+213 between May 27, 2014 (MJD 56804) and July 2, 2014 (MJD 56840) with 
27.8~hours of total live time. Observations took place at elevations between 63$^{\circ}$ and 80$^{\circ}$, 
leading to $\sim$18$\sigma$ source detection above 200~GeV.



%


\section{Results}
\begin{figure}[!ht]
\includegraphics[width=80mm]{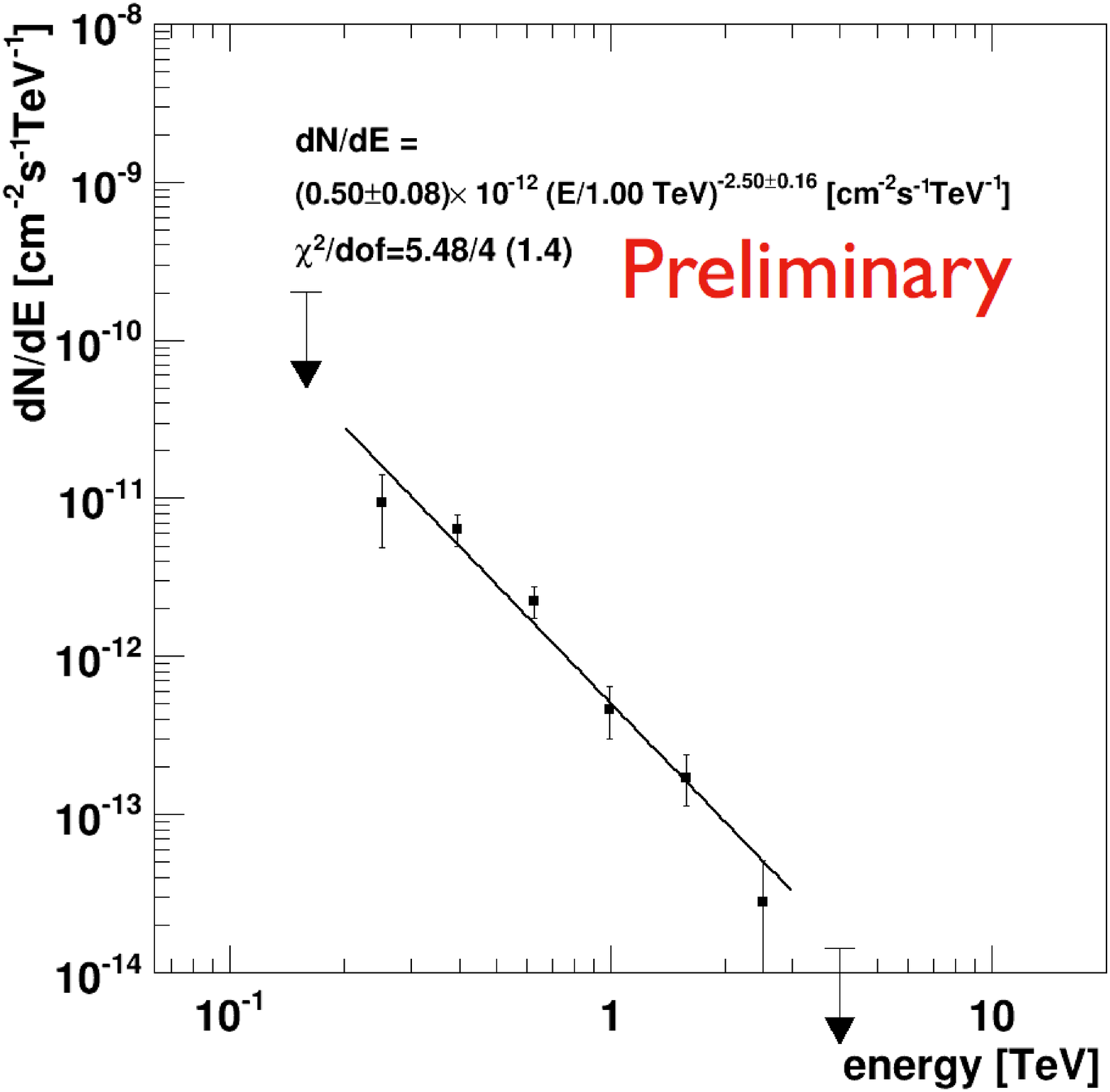}
\caption{Differential energy spectrum of HESS J1943+213 with VERITAS, between 200~GeV and 2~TeV, with results from a fit to a power-law function.}
\label{spectrum}
\end{figure}

Figure~\ref{sigmap} shows the significance sky map near HESS J1943+213, including the detected source. The source location is 
consistent with the catalog position of HESS J1943+213.
A preliminary VERITAS differential energy spectrum of the source is shows in Figure~\ref{spectrum}. The high elevation observations allow for a lower energy threshold and lead to a spectrum that extends down to 200~GeV, compared with 470~GeV from H.E.S.S. observations. The source spectrum is fit by a power-law function with an index of $\Gamma$=2.5 $\pm$ 0.16 in the energy range 200~GeV--2~TeV. The VERITAS spectrum of HESS J1943+213 appears harder than the spectrum from H.E.S.S. ($\Gamma$=3.1 $\pm$ 0.3), though a more rigorous comparison is necessary, as the energy ranges differ between the two detections. 

The flux of (1.30 $\pm$ 0.20) $\times$ 10$^{-12}$ cm$^{-2}$ s$^{-1}$ measured with VERITAS above 470 GeV is consistent with a flux of (1.25 $\pm$ 0.20) $\times$ 10$^{-12}$ cm$^{-2}$ s$^{-1}$ from the H.E.S.S. detection. Additionally, no flux variability is seen within VERITAS observations. 

As VERITAS is able to observe HESS J1943+213 at a much higher elevation than H.E.S.S., the detection rate of the source with VERITAS is 3.4$\sigma$/$\sqrt{\textrm{hour}}$, compared with 1.8$\sigma$/$\sqrt{\textrm{hour}}$ with H.E.S.S., allowing VERITAS to test for variability on a factor of four shorter timescales.

\section{Discussion and Outlook}

The agreement between fluxes measured approximately five years apart with VERITAS and H.E.S.S. and a lack of a variability detection in other energy bands is surprising if HESS J1943+213 is a blazar. Blazars are known to vary at all energies and at a wide range of timescales~\cite{boettcher}. Although the stable flux of the source observed to date does not rule out the blazar hypothesis, continued non-detections of variability are becoming a growing challenge for this scenario. 

VERITAS can probe timescales that are a factor of four shorter than those available to H.E.S.S and therefore represents the best VHE dataset available for searches of flux and spectral variability from this source. Advanced analysis techniques will provide an even higher sensitivity and allow for an additional factor of two improvement in the minimum variability timescale that can be tested.

The VERITAS-measured spectrum of HESS J1943+213, albeit preliminary, exhibits a harder index than the H.E.S.S. spectrum. The soft spectral index from H.E.S.S. constitutes one of the key pieces of evidence against the PWN hypothesis. Thus, in conjunction with the lack of detectible variability in the VHE regime, the harder VERITAS spectrum may be counted in favor of the PWN hypothesis.

VERITAS will continue observations of HESS J1943+213 and will monitor the source for potential variability. In addition, the upcoming PASS 8 Fermi-LAT data will allow for an improved detection and a spectrum of the source in GeV gamma rays, providing a significantly better handle on the gamma-ray peak of the HESS J1943+213 spectral energy distribution. Multi-wavelength studies of the source, including studies of its spectral energy distribution will be essential for definitively identifying the source class.

%






\end{document}